\documentclass[showpacs,preprint,showpacs,preprintnumbers,amsmath,amssymb]{revtex4}
\usepackage{graphicx} 
\usepackage{color}
\usepackage{amssymb,amsfonts,amsmath}
\usepackage{pdfsync}
\usepackage{caption}
\usepackage{tgtermes}
\usepackage{epsfig}
\def \BEA {\begin{eqnarray}}
\def \EEA {\end{eqnarray}}

\begin{document}

\title{Derivation of the Four-Wave Kinetic Equation in Action-Angle Variables} 

\author{M. Onorato$^{1,2}$}
\author{G. Dematteis$^{1}$}
\affiliation{
$^1$Dip. di Fisica, Universit\`{a} di Torino, Via P. Giuria, 1 - Torino, 10125, Italy; \\
$^2$ INFN, Sezione di Torino, Via P. Giuria, 1 - Torino, 10125, Italy
}

\begin{abstract}
Starting from the action-angle variables and using a standard asymptotic expansion, here we present a new derivation of the Wave Kinetic Equation for resonant process of the type 2 $\leftrightarrow$ 2. Despite not offering new physical results and despite not being more rigorous than others, our procedure has the merit of being  straightforward; it allows for a direct control of the random phase and random amplitude hypothesis of the initial wave field. We show that the  Wave Kinetic Equation can be derived assuming only initial random phases. The random amplitude approximation has to be taken only at the end, after taking the weak nonlinearity and large box limits. This is because the $\delta$-function over frequencies  contains the amplitude-dependent nonlinear correction which should be dropped before the random amplitude approximation applies. If $\epsilon$ is the small parameter in front of the anharmonic part of the Hamiltonian, the time scale associated with the Wave Kinetic equation is shown to be  $1/\epsilon^2$. We give evidence that  random phase and amplitude hypotheses persist up to a time of the order $1/\epsilon$. 
 \end{abstract}
\maketitle
\section{Introduction}
The Wave Kinetic (WK) equation is an important tool for describing  systems of 
interacting waves \cite{nazarenko2011wave}; it finds applications in many fields of physics such 
as gravity, capillary and internal waves, plasma waves, Bose Einstein condensation, elastic plate waves, etc.
At the moment there is no derivation of the WK equation that can be considered
as rigorous in a mathematical sense. However, physicists have attempted different roads:
two are the main procedures. The first one  
is the direct derivation of the WK equation by performing statistical averaging over 
the equations of motion \cite{hasselmann62,falkovich1992kolmogorov,newell1968closure},
 the other is through the calculation of the first moment of the equation for the
 probability density function for  the amplitudes and phases \cite{choi2004probability,choi2005joint,eyink2012kinetic,chibbaro20184}.
 Each derivation has its own strengths; at the same time, none of them 
 seems to be adequately rigorous. In particular, while the first kind of derivation assumes the {\it propagation of chaos} to justify the validity of the equation at positive times, the second tries to go beyond and to prove that independent uniform phases and independent amplitudes (RPA assumption) of the initial field are sufficient to preserve the RPA hypothesis at later times. Though, none of the mentioned derivations has made an attempt (possible, in principle) to control rigorously the remainder terms of the small-$\epsilon$ perturbation expansion. The hope is that these higher order terms give a negligible contribution in the small-$\epsilon$ limit, in analogy to what Lanford has proved in the low-density limit for gases. 
Some attempts to make a rigorous derivation can be found in \cite{lukkarinen2011weakly,kuksin2015derivation,buckmaster2019onset}.
 
 Our derivation does not pretend to be more rigorous than the 
 existing ones;
 however, according to us, it has the merit of being straightforward. It is based on a 
 direct expansion of the variables angle and action in powers of the small parameter
 in front of the interaction Hamiltonian.
  Because we use angle-action variables we are able control in a clear way the two different procedures of  averaging, i.e. over initial angles and over initial actions.
  With respect to other derivations where an auxiliary intermediate time scale is introduced, see \cite{nazarenko2011wave}, in our derivation such time scale arises naturally from the expansion. 
It is sufficient to average over angles the equation for the actions to show  that the time scale of the evolution of the action variable scales like $1/\epsilon^2$, where $\epsilon$ is the perturbation parameter. The dynamics on a shorter nonlinear time scale $1/\epsilon$ simply averages out. 
We show that the phases and the actions remain uncorrelated over a time scale of $1/\epsilon$.
We hope that our approach could pave the way for a mathemathical derivation of the equation.
We also hope that students entering into the field may benefit from it.

\section{The Hamiltonian model}
We consider a Hamiltonian system with quartic nonlinearity in the Hamiltonian which 
allows, in the thermodynamic limit, for resonances of the type 2$\leftrightarrow$2.

The starting physical space is defined as $\Lambda=[0,L]\in\mathbb{R}^d$. Its dual is the infinite discrete Fourier space $\Lambda^*=\frac{2\pi}{L}\mathbb{Z}^d$.
Some shorthand notation:
$$
\sum_{1234}:=\sum_{k_1,k_2,k_3,k_4}, \text{with }k_i\in\Lambda^*, \qquad \delta^{12}_{34} := \delta_{k_1+k_2,k_3+k_4} \text{(Kr\"onecker delta)},
$$
$$
y_1:=y(k_1), \qquad \Delta y_{12}^{34}:=y_1+y_2-y_3-y_4\,.
$$
Summations go from -$\infty$ to +$\infty$.

In  normal variable $a_k$ the Hamiltonian takes the following form:
\begin{equation}
	\mathcal H = \sum_{k_1}\omega_1 |a_{1}|^2 +\frac{\epsilon}{2} \sum_{1,2,3,4} T_{1234}a_1^*a_2^*a_3a_4 \delta_{12}^{34} \,,
\end{equation}
 where $\omega_k=\omega_{-k}\ge 0$ is the dispersion relation and $\epsilon\ll1$  is the weak nonlinearity parameter.
 For $T_{1234}=const$ and $\omega(k)=k^2$, the Hamiltonian is the one for the Nonlinear Schr\"odinger equation.
Performing the following transformation
\begin{equation}
a_k=\sqrt{I_k}\exp(-i \theta_k),
\end{equation}
the Hamiltonian can be written in canonical action-angle variable, $\{I_k, \theta_k\}$,  as
\begin{equation}
	\mathcal H = \sum_{k_1}\omega_1 I_1 +\frac{\epsilon}{2} \sum_{1,2,3,4} T_{1234}\sqrt{I_1I_2I_3I_4} \cos(\Delta{\theta}_{12}^{34})\delta_{12}^{34}.
\end{equation}
For issues related to what follows, a simplification in the procedure can be achieved if
the diagonal terms are extracted from the sums, so that the Hamiltonian takes the following form:
\begin{equation}
	\mathcal H = \sum_{k_1}\Omega_1 I_1 +\frac{\epsilon}{2} \sum_{1,2,3,4}'T_{1234}\sqrt{I_1I_2I_3I_4} \cos(\Delta{\theta}_{12}^{34})\delta_{12}^{34},
\end{equation}
where the sum $\sum'_{1234}$ excludes all cases for which either $k_1= k_3$ and $k_2= k_4$, or $k_1= k_4$ and $k_2= k_3$, or $k_1=k_2=k_3=k_4$ which are the trivial resonances  and 
the renormalized dispersion relation is introduced:
\begin{equation}
\Omega_k=\omega_k+2\epsilon\sum_{k'}T_{kk'kk'}I_{k'}-\epsilon T_{kkkk}I_k.
\end{equation}

Hamilton's equations  take the form:
\begin{equation}\label{eq:4}
\left\{\begin{aligned}
&\frac{d I_k}{d t} = -\frac{\partial \mathcal H}{\partial \theta_k} =2 \epsilon\sum_{234}' T_{k234}\sqrt{I_kI_2I_3I_4} \sin(\Delta \theta_{k2}^{34})\delta_{k2}^{34} \\
&\frac{d \theta_k}{d t} = \frac{\partial \mathcal H}{\partial I_k} = \Omega_k + \epsilon\sum_{234}' T_{k234}\sqrt{\frac{I_2I_3I_4}{I_k}} \cos(\Delta \theta_{k2}^{34})\delta_{k2}^{34}
\end{aligned}\right.
\end{equation}
with initial data:
\begin{equation}
I_k(t=0)={\bar I}_k,\;\;\;\;\;
\theta_k(t=0)={\bar \theta}_k
\end{equation}

%
\section{The  $\epsilon$-expansion}
We perform the small-$\epsilon$ power expansion
\begin{equation}
\begin{split}
\label{eq:expansion}
&I_k(t) = I_k^{(0)}(t)+\epsilon I_k^{(1)}(t) + \epsilon^2 I_k^{(2)}(t)+\mathcal{O}(\epsilon^3)\\
&\theta_k(t) = \theta_k^{(0)}(t) + \epsilon \theta_k^{(1)}(t)+ \epsilon^2 \theta_k^{(2)}(t)+\mathcal{O}(\epsilon^3)
\end{split}
\end{equation}
and plug into \eqref{eq:4} to obtain, order by order,
\begin{itemize}
\item$\epsilon^0$: \\
Linear evolution where only the fast angle oscillations are at play,
\begin{equation}\label{eq:7}
\begin{array}{cc}
\left\{\begin{aligned}
&\frac{dI_1^{(0)}}{dt} =  0 \\
&\frac{d\theta_1^{(0)}}{dt} = \bar\Omega_1
\end{aligned}\right.\qquad
\Rightarrow\qquad
\left\{\begin{aligned}
& I_1^{(0)}(t) =  {\bar I}_k =const \\
&\theta_1^{(0)} = {\bar \theta}_k+\bar\Omega_1 t\;\;\;\; \mod 2\pi
\end{aligned}\right.
\end{array}
\end{equation}
Here $\bar\Omega_k=\omega_k+\epsilon2\sum_{k_2}T_{kk_2kk_2}\bar{I}_2-\epsilon T_{kkkk}\bar{I}_k.$
While the angles evolve on the linear time scale, the variations for the actions require a higher order dynamics in $\epsilon$. Note also that the linear time scale $1/\bar\Omega_k$ is 
$k$ dependent; this implies that for example for dispersion relations for which 
$\bar\Omega_k \rightarrow 0$ for $k \rightarrow 0$ , then the linear time scale may become extremely large.

\item$\epsilon^1$: \\
\begin{equation}\label{eq:8}
\left\{\begin{aligned}
&\frac{d I_1^{(1)}}{dt} = 2 \sum'_{234} T_{1234}\sqrt{\bar{I}_1\bar{I}_2\bar{I}_3\bar{I}_4} \sin({\Delta\theta^{(0)}}_{12}^{34})\delta_{12}^{34} \\
&\frac{d\theta_1^{(1)}}{dt} =\sum'_{234} T_{1234}\sqrt{\frac{\bar{I}_2\bar{I}_3\bar{I}_4}{\bar{I}_1}} \cos({\Delta\theta^{(0)}}_{12}^{34})\delta_{12}^{34}
\end{aligned}\right.
\end{equation}
Integrating in time  yields
\begin{equation}\label{eq:9}
\left\{\begin{aligned}
& I_1^{(1)} = 2 \sum'_{234} T_{1234}\sqrt{\bar{I}_1\bar{I}_2\bar{I}_3\bar{I}_4} \frac{\cos\left(\Delta\bar{\theta}_{12}^{34}\right)-\cos\left(\Delta\bar{\theta}_{12}^{34}+\Delta \bar\Omega_{12}^{34}t\right)}{\Delta \bar\Omega_{12}^{34}}\delta_{12}^{34} \\
& \theta_1^{(1)} =\sum'_{234} T_{1234}\sqrt{\frac{\bar{I}_2\bar{I}_3\bar{I}_4}{\bar{I}_1}} \frac{\sin\left(\Delta\bar{\theta}_{12}^{34}+\Delta \bar\Omega_{12}^{34}t\right) - \sin\left(\Delta\bar{\theta}_{12}^{34}\right)}{\Delta \bar\Omega_{12}^{34}}\delta_{12}^{34}
\end{aligned}\right.
\end{equation}
\item$\epsilon^2$: \\
\begin{equation}\label{eq:10}
\begin{aligned}
\frac{ dI_1^{(2)}}{dt}& = 2 \sum'_{234} T_{1234}\sqrt{\bar{I}_1\bar{I}_2\bar{I}_3\bar{I}_4} \Bigg[\frac12\left( \frac{I_1^{(1)}}{\bar{I}_1}+\frac{I_2^{(1)}}{\bar{I}_2}+\frac{I_3^{(1)}}{\bar{I}_3}+\frac{I_4^{(1)}}{\bar{I}_4}\right)\sin\left(\Delta\bar{\theta}_{12}^{34}+\Delta \bar\Omega_{12}^{34}t\right) +
\\&+ {\Delta\theta^{(1)}}_{12}^{34} \cos\left(\Delta\bar{\theta}_{12}^{34}+\Delta \bar\Omega_{12}^{34}t\right)\Bigg]  \delta_{12}^{34}
\end{aligned}
\end{equation}
which substituting the expressions in \eqref{eq:9} leads, after some algebra and 
the use of trigonometric identities,  to this compact form
\begin{equation}\label{eq:secondorder0}
\begin{aligned}
&\frac{d I_1^{(2)}}{dt} = 2\sum_{m=1}^4 \sum'_{234} \sum'_{567} T_{1234}T_{m567}
 \frac{1}{\sqrt{\bar{I}_m}} \sqrt{\bar{I}_1\bar{I}_2\bar{I}_3\bar{I}_4\bar{I}_5\bar{I}_6\bar{I}_7}
\times \\
\Bigg[ 
&
\frac{\sin[\sigma_m\Delta\bar{\theta}_{m5}^{67}-\Delta\bar{\theta}_{12}^{34}+
(\sigma_m\Delta \bar\Omega_{m5}^{67} -\Delta \bar\Omega_{12}^{34})t]+
\sin[\Delta\bar{\theta}_{12}^{34}-\sigma_m \Delta\bar{\theta}_{m5}^{67}+\Delta \bar\Omega_{12}^{34}t]}{ \Delta \bar\Omega_{m5}^{67}}\Bigg]
\end{aligned}
\end{equation}
where $\sigma=(+1,+1,-1,-1)$.
The evolution equation for $\theta_k^{(2)}$ is not needed.

\end{itemize}
%

For the evolution of the action variable we have thus obtained:
\begin{equation}\label{eq:actionevol}
\frac{d I_k}{dt} = \epsilon \frac{d I_k^{(1)}}{dt} + \epsilon^2 \frac{d I_k^{(2)}}{dt}+\mathcal{O}(\epsilon^3)\,,
\end{equation}
where the terms on the right hand side are given respectively by \eqref{eq:8} and \eqref{eq:secondorder0}.

\section{Averaging over  over initial angles: the discrete wave kinetic equation}
Assuming that the phases are random variables, distributed uniformly in the interval 
$[0,2\pi)$, we define the procedure of averaging over initial phases, $\bar{\theta_k}$,
 of the observable $g_k$ as:
\begin{equation}
\langle g_k \rangle_{\bar{\theta}}=\frac{1}{2\pi}\int_0^{2\pi} g_k d\bar{\theta_k}
\end{equation}

We are interested in the following: 
\begin{equation}\label{eq:12}
\langle\frac{d I_k}{dt}\rangle_{\bar \theta} = \epsilon \langle \frac{d I_k^{(1)}}{dt} \rangle_{\bar \theta}+ \epsilon^2  \langle \frac{d I_k^{(2)}}{dt}\rangle_{\bar \theta} +\mathcal{O}(\epsilon^3)\,,
\end{equation}
Two time scales appears in the equation (\ref{eq:12}); however, as it will be clear soon, the procedure of  averaging over the initial phases cancels the shortest time scale.

\begin{itemize}
\item $\epsilon$\\
\begin{equation}\label{eq:firstorder}
\frac{d \langle I_1^{(1)}\rangle_{\bar \theta}}{dt} = 2\langle \sum'_{234} T_{1234}\sqrt{\bar{I}_1\bar{I}_2\bar{I}_3\bar{I}_4}  \sin({\Delta\theta^{(0)}}_{12}^{34})\delta_{12}^{34} 
\rangle_{\bar \theta}\\
\end{equation}
Because of the presence of the  $\sin$ function, it is straightforward to show that the contribution is 0:
\begin{equation}\label{eq:firstorder}
\frac{d \langle I_1^{(1)}\rangle_{\bar \theta}}{dt} =0
\end{equation}
This implies that the change of the actions depends, after phase averaging, on higher order 
contributions.
\item $\epsilon^2$\\
Using some trigonometric identities, eq. (\ref{eq:secondorder0}) can be rewritten
\begin{equation}\label{eq:secondorder}
\begin{aligned}
& \frac{d I_1^{(2)}}{dt}= 2  \sum_{m=1}^4 \sum'_{234} \sum'_{567} T_{1234}T_{m567}
 \frac{1}{\sqrt{\bar{I}_m}} \sqrt{\bar{I}_1\bar{I}_2\bar{I}_3\bar{I}_4\bar{I}_5\bar{I}_6\bar{I}_7}
\times \\
\Bigg[ 
&
\frac{\cos(\sigma_m\Delta\bar{\theta}_{m5}^{67}-\Delta\bar{\theta}_{12}^{34})[\sin((\sigma_m\Delta \bar\Omega_{m5}^{67} -\Delta \bar\Omega_{12}^{34})t)+\sin(\Delta \bar\Omega_{12}^{34}t)]}{\Delta \bar\Omega_{m5}^{67}}+
\\
&+
\frac{\sin\left(\sigma_m\Delta\bar{\theta}_{m5}^{67}-\Delta\bar{\theta}_{12}^{34})[\cos((\sigma_m\Delta \bar\Omega_{m5}^{67} -\Delta \bar\Omega_{12}^{34})t \right)-
\cos(\Delta \bar\Omega_{12}^{34}t)]}{\Delta \bar\Omega_{m5}^{67}} 
\Bigg] \delta_{12}^{34}\delta_{m5}^{67}.
\end{aligned}
\end{equation}
The r.h.s. of equation (\ref{eq:secondorder}) is composed by two terms, each written as  
product of two trigonometric functions: the first factor has an argument which includes
only initial phases; this form is suitable for the phase averaging  procedure.
The first term can be written in general as
\begin{equation}\label{eq:secondorder1}
\begin{aligned}
\langle\sum_{m=1}^4 \sum'_{234} \sum'_{567} F_{1234}G_{m567}
\cos(\sigma_m\Delta\bar{\theta}_{m5}^{67}-\Delta\bar{\theta}_{12}^{34})
\rangle_{\bar \theta}
\end{aligned}
\end{equation}
with obvious meaning of  $F_{1234}$ and $G_{m567}$.
We show the calculation of phase averaging for $m=1$.\\
The sum can be always split as follows:
\begin{equation}
\begin{split}
&\langle \sum_{k_2,k_3,k_4}'\sum_{k_5,k_6,k_7}'F_{1234}G_{1567}\cos(\Delta\bar{\theta}_{256}^{345})\rangle_{\bar\theta}=\sum_{k_2,k_3,k_4}'F_{1234}(G_{1234} + G_{1243}) + 
\\ 
&+\sum_{k_2,k_3,k_4}'\sum'_{\substack{k_5\neq k_2\\k_6\neq k_3,k_6\neq k_4\\k_7\neq k_3,k_7\neq k_4}}
F_{1234}G_{1567}\langle  \cos(\Delta\bar{\theta}_{256}^{345}) \rangle_{\bar{\theta}}= 2\sum_{k_2,k_3,k_4}'F_{1234}G_{1234}
\end{split}
\end{equation}
\begin{equation}
\begin{split}
&\langle \sum_{k_2,k_3,k_4}'\sum_{k_5,k_6,k_7}'F_{1234}G_{1567}\cos(\Delta\bar{\theta}_{267}^{345})\rangle_{\bar\theta}=\sum_{k_2,k_3,k_4}'F_{1234}(G_{1234} + G_{1243}) + 
\\ 
&+\sum_{k_2,k_3,k_4}'\sum'_{\substack{k_5\neq k_2\\k_6\neq k_3,k_6\neq k_4\\k_7\neq k_3,k_7\neq k_4}}
F_{1234}G_{1567}\langle  \cos(\Delta\bar{\theta}_{267}^{345}) \rangle_{\bar{\theta}}= 2\sum_{k_2,k_3,k_4}'F_{1234}G_{1234}
\end{split}
\end{equation}
where the second term has been dropped because the argument of the cosine is a sum of random variables that are uniformly distributed over $[0,2\pi)$ and therefore its distribution is also uniform over $[0,2\pi)$ and its average is zero.
The calculation can be done for $m=2,3,4$ in a similar fashion.

Following the same procedure, it is straightforward to show that in eq. (\ref{eq:secondorder}) the $\sin$ term containing the initial phases averages out.
The final evolution equation for $I_k^{(2)}$ reads
\begin{equation}\label{eq:15}
\langle \frac{dI_1^{(2)}}{dt}\rangle_{\bar\theta} = 4 \sum_{234}T_{1234}^2 \bar{I}_1\bar{I}_2\bar{I}_3\bar{I}_4 \left( \frac{1}{\bar{I}_1}+\frac{1}{\bar{I}_2}-\frac{1}{\bar{I}_3}-\frac{1}{\bar{I}_4}\right) \frac{\sin\left(\Delta \bar\Omega_{12}^{34}t\right)}{\Delta \bar\Omega_{12}^{34}} \delta_{12}^{34}
\end{equation}
Note that there is no need to use the reduced sum $\sum'$ symbol, because the extra terms in the standard sum give a zero contribution (exact cancellations due to the two ``$+$'' signs and two ``$-$'' signs in the term in brackets). \\
\end{itemize}

Inserting eq. (\ref{eq:15}) in eq. (\ref{eq:12}), it gives
\begin{equation}\label{eq:discretekinetic}
 \frac{d I_1 }{dt} = \epsilon^2 4 \sum_{234}T_{1234}^2 \bar{I}_1\bar{I}_2\bar{I}_3\bar{I}_4 \left( \frac{1}{\bar{I}_1}+\frac{1}{\bar{I}_2}-\frac{1}{\bar{I}_3}-\frac{1}{\bar{I}_4}\right) \frac{\sin\left(\Delta \bar\Omega_{12}^{34}t\right)}{\Delta \bar\Omega_{12}^{34}} \delta_{12}^{34}
\end{equation}

 If we define the nonlinear time $\tau=\epsilon^2 t$, then the equation reads
\begin{equation}\label{eq:discretekinetic}
 \frac{d I_1}{d\tau} = 4 \sum_{234}T_{1234}^2 \bar{I}_1\bar{I}_2\bar{I}_3\bar{I}_4 \left( \frac{1}{\bar{I}_1}+\frac{1}{\bar{I}_2}-\frac{1}{\bar{I}_3}-\frac{1}{\bar{I}_4}\right) \frac{\sin\left(\Delta \bar\Omega_{12}^{34}\tau/\epsilon^2\right)}{\Delta \bar\Omega_{12}^{34}} \delta_{12}^{34}.
\end{equation}
{If  $T_{kk'kk'}=const$ and $\sum_k I_k$ is conserved (this property is shared by the Nonlinear Schr\"odinger equation), then the nonlinear frequency shift contribution is identically zero and, in the limit of 
$\epsilon\rightarrow 0$, eq. (\ref{eq:discretekinetic}) becomes
\begin{equation}\label{eq:discretekinetic2}
 \frac{d I_1}{d\tau} = 4 \pi\sum_{234}T_{1234}^2 \bar{I}_1\bar{I}_2\bar{I}_3\bar{I}_4 \left( \frac{1}{\bar{I}_1}+\frac{1}{\bar{I}_2}-\frac{1}{\bar{I}_3}-\frac{1}{\bar{I}_4}\right) 
 \delta({\Delta \omega_{12}^{34}} )\delta_{12}^{34},
\end{equation}
where we have used the property that
\begin{equation}\label{eq:resonance}
\lim_{\epsilon\to 0} \frac{\sin\left(\Delta \omega_{12}^{34}\tau/\epsilon^2\right)}{\Delta \omega_{12}^{34}} =\pi \delta(\Delta \omega_{12}^{34}).
\end{equation}

{\it Remarks}\\
$\bullet$ The $\delta(\Delta\omega_{12}^{34})$ above is dimensionally a Dirac delta, coming from the limit relationship (\ref{eq:resonance}). This is not rigorous and in principle even not meaningful, being the argument of the $\delta$ not a continuous function. Though, one can argue that the values taken by $\Delta\omega_{12}^{34}$ can become extremely dense around  $\Delta\omega_{12}^{34}=0$, which can be thought of as the summation tending to an integral.\\
$\bullet$ The time scale for the evolution of the action is $1/\epsilon^2$.\\
$\bullet$ For the validity of the expansion, such time scale should always  be much larger than the linear time scale given by $1/\omega_k$ for all values of $k$.\\
$\bullet$ In the r.h.s. only the initial actions are included.\\
$\bullet$ The highest order contribution to the evolution of $I_k$ at order $\epsilon^2$ from the angle dynamics comes from $\theta_k^{(1)}$. No contribution from  $\theta_k^{(2)}$ enters.  \\
$\bullet$ No assumptions on the statistics of initial actions has been made.\\
$\bullet$ The equation (\ref{eq:discretekinetic2}) is meaningful only if the dispersion relation allows for connected exact resonances  on a regular discrete grid.
}

\section{The thermodynamic limit: the standard wave kinetic equation}
The physical space over which we have worked is defined as $\Lambda=[0,L]\in\mathbb{R}^d$.
In the thermodynamic limit one is interested in looking  at the limit $L \rightarrow \infty$. As
 this limit is taken, the spacing between Fourier modes $\Delta k=2\pi/L$ becomes smaller and smaller in such  a way that wave numbers $k \in\mathbb{R}^d$. In this limit a resonant manifold, that could be empty in the case of regular discrete grid, may appear. Therefore, the starting point
for the derivation should be equation (\ref{eq:discretekinetic}), where still the Dirac Delta function over frequencies is not introduced. The thermodynamic limit
($\Delta k\rightarrow 0$ or $L\rightarrow \infty$) is taken using the following rules:
  
$\bullet$ We define the action density as:
 \begin{equation}
{\rm I}_k={\rm I}(k,t):= \frac{ I_k }{\Delta k^d} ,
 \end{equation}
 where ${\rm I}(k,t)$ is a continuous function of $k \in\mathbb{R}^d$
 
 $\bullet$ Sums become integrals as follows:
 \begin{equation}
\sum_k\rightarrow \int \frac{1}{\Delta k^d}d k
 \end{equation}
 
  $\bullet$ The Kronecker Delta $\delta^{(K)}$ becomes a Dirac Delta $\delta^{(D)}$
 \begin{equation}
\delta^{(K)} \rightarrow \Delta k^d \delta^{(D)}
 \end{equation}

Introducing the above substitutions in \eqref{eq:discretekinetic}, we get:
\begin{equation}\label{eq:discretekinetic_avI}
 \frac{d{\rm I}_1}{d \tau} = 4 \int dk_2 dk_3 dk_4 T_{1234}^2
  \bar{{\rm I}}_1  \bar{{\rm I}}_2 \bar{{\rm I}}_3  \bar{{\rm I}}_4 
  \left( \frac{1}{ \bar{{\rm I}}_1}+\frac{1}{\bar{{\rm I}}_2}- \frac{1}{ \bar{{\rm I}}_3}-\frac{1}{ \bar{{\rm I}}_4}\right) 
\frac{\sin\left(\Delta \bar\Omega_{12}^{34}\tau/\epsilon^2\right)}{\Delta \bar\Omega_{12}^{34}}\delta_{12}^{34},
\end{equation}
{
where we need to take the limit for $\Delta k\rightarrow 0$  of
\begin{equation}
\lim_{\Delta k \to 0}\frac{\sin\left[(\omega_k+\Delta k^d\epsilon2\sum_{k_2}T_{kk_2kk_2}\bar{\rm I}_2-\epsilon \Delta k^d T_{kkkk}\bar{\rm I}_k)\tau/\epsilon^2\right]}{(\omega_k+\Delta k^d\epsilon2\sum_{k_2}T_{kk_2kk_2}\bar{\rm I}_2-\Delta k^d\epsilon T_{kkkk}\bar{\rm I}_k )}=
\frac{\sin\left(\omega_k\tau/\epsilon^2\right)}{\omega_k}
\end{equation}
The last equality is valid only if we assume that 
\begin{equation}
\lim_{\substack{\Delta k \to 0 \\ \epsilon \to 0}}\frac{\Delta k^d}{\epsilon}=0
\end{equation}

\subsection{The weakly nonlinear limit}
By taking the small amplitude  limit $\epsilon\rightarrow 0$, one gets
\begin{equation}\label{eq:discretekinetic_avI}
 \frac{d{\rm I}_1}{d \tau} = 4 \pi \int dk_2 dk_3 dk_4 T_{1234}^2
  \bar{{\rm I}}_1  \bar{{\rm I}}_2 \bar{{\rm I}}_3  \bar{{\rm I}}_4 
  \left( \frac{1}{ \bar{{\rm I}}_1}+\frac{1}{\bar{{\rm I}}_2}- \frac{1}{ \bar{{\rm I}}_3}-\frac{1}{ \bar{{\rm I}}_4}\right) 
 \delta({\Delta \omega_{12}^{34}} )\delta_{12}^{34},
\end{equation}

\subsection{The assumption of initial uncorrelated random actions}
We now assume that ${\rm I}_k$ is a stochastic variable whose expectation value made with respect to the distribution of the  initial actions is given by
\begin{equation}
n(k,t)=\langle {\rm I}(k,t)\rangle_{\bar{\rm I}_k};
\end{equation} 
the equation above defines the spectral action density $n(k,t)$ or more simply the action spectrum.
We  assume that actions are uncorrelated in wave numbers,  so that
\begin{equation}
\langle \bar{\rm I}_i\bar{\rm I}_j\bar{\rm I}_k\rangle_{\bar {\rm I}}=
\langle \bar{\rm I}_i\rangle_{\bar{\rm I}}\langle \bar{\rm I}_j \rangle_{\bar{\rm I  }}\langle 
\bar{\rm I}_k\rangle_{\bar {\rm I}}=n_i n_j n_k, \;i\neq j\neq k;
\end{equation}
therefore, the equation for the spectrum becomes
\begin{equation}\label{eq:discretekinetic_avI}
 \frac{d n_1}{d\tau} = 4 \pi \int dk_2 dk_3 dk_4 T_{1234}^2
\bar n_1 \bar n_2 \bar n_3 \bar n_4 \left( \frac{1}{\bar n_1}+\frac{1}{\bar n_2}-\frac{1}{\bar n_3}-\frac{1}{\bar n_4}\right) 
 \delta({\Delta \omega_{12}^{34}} )\delta_{12}^{34},
\end{equation}

Once more,
in the right hand side of the equation only initial data for $n_k$ are included. Thus, strictly speaking its
validity is at time $t=0$.
A usual but somehow unjustified further step consists in putting also on the right hand side the spectral action density $n_k(t)$ instead of $\bar{n}_k=n_k(t=0)$ to get:

\begin{equation}\label{eq:discretekinetic_avI1}
 \frac{dn_1}{d \tau} = 4 \pi \int dk_2 dk_3 dk_4 T_{1234}^2
 n_1n_2n_3n_4
   \left( \frac{1}{ n_1}+\frac{1}{n_2}- \frac{1}{ n_3}-\frac{1}{n_4}\right) 
 \delta({\Delta \omega_{12}^{34}} )\delta_{12}^{34},
\end{equation}
This is the celebrated Wave Kinetic Equation.

\section{Persistence of the initial statistics}
At time $t=0$, the statistical properties of the actions and angles is prescribed:
in particular, the initial angles are distributed uniformly in $[0, 2\pi)$ and are uncorrelated; the initial actions 
are considered as uncorrelated, i.e. $\langle \bar{I}_i \bar{I}_j \rangle= 
\langle \bar{I}_i\rangle \langle \bar{I}_j\rangle$. We would like to know if these conditions persist
up to the time scale of collision.

\subsection{The actions}
Using the expansion in (\ref{eq:expansion}), we plug it in the second order correlation function:
\begin{equation}
\begin{split}
&\langle I_1(t) I_2(t) \rangle -\langle I_1(t)\rangle \langle I_2(t) \rangle
= \langle \bar{I}_1 \bar{I}_2 \rangle-\langle \bar{I}_1\rangle \langle \bar{I}_2 \rangle+\\
&+\epsilon
\left(\langle \bar{I}_1 I_2^{(1)}\rangle -\langle \bar{I}_1\rangle \langle I_2^{(1)} \rangle+
\langle \bar{I}_2 I_1^{(1)}\rangle  -\langle \bar{I}_2\rangle \langle I_1^{(1)} \rangle \right)+
\mathcal{O}(\epsilon^2)
\end{split}
\end{equation}
where $ I_k^{(1)}$ is given by eq. (\ref{eq:9}). 
The first terms on the right hand side vanish because  the initial data satisfy
$\langle \bar{I}_1 \bar{I}_2 \rangle=\langle \bar{I}_1\rangle \langle \bar{I}_2 \rangle$.
Because $\bar{I}_1$ does not depend on angles, then  $\langle \bar{I}_1 I_2^{(1)}\rangle=\langle \bar{I}_1 \langle I_2^{(1)} \rangle_{\bar{\theta}} \rangle_{\bar{I}}=0$; this is  because $ \langle I_2^{(1)} \rangle_{\bar{\theta}}=0$. Similarly, all terms in $\epsilon$ vanish. Therefore, for time of the order of $1/\epsilon$, the actions remain uncorrelated. 
\begin{equation}
\begin{split}
&\langle I_1(t) I_2(t) \rangle =\langle I_i(t)\rangle \langle I_j(t) \rangle
+
\mathcal{O}(\epsilon^2)
\end{split}
\end{equation}

The same procedure can be used to study the correlation of $n$ actions, for $n$ arbitrarily large.


\subsection{Angles}
For angles at leading order one should remember that 
\begin{equation}
\theta^{(0)}_k(t)=\bar{\theta}_k+\bar\Omega_k t\;\;\;  \mod 2\pi
\end{equation}
This is a well known equation in the theory of dynamical systems. In particular, if waves are dispersive, then the system is 
ergodic (for irrational values of $\bar\Omega_k$) and the distribution of $\theta^{(0)}_k$ is uniform as for $\bar{\theta}_k$
and  $\theta^{(0)}_k(t)$ are uncorrelated for all times at leading order.
This is an important result, telling us that averages can be computed over 
$\theta^{(0)}_k(t)$. 
We can show that the angles remain uncorrelated for the time scale $1/\epsilon$
\begin{equation}
\begin{split}
&\langle \theta_1(t) \theta_2(t) \rangle -\langle \theta_1(t)\rangle \langle \theta_2(t) \rangle
=
 \langle \theta_1^{(0)}(t) {\theta}_2^{(0)}(t) \rangle-\langle \theta_1^{(0)}(t)\rangle \langle {\theta}_2^{(0)}(t) \rangle+\\
&+\epsilon
\left(\langle {\theta}_1^{(0)} (t)\theta_2^{(1)}(t)\rangle -\langle {\theta}_1^{(0)}(t)\rangle \langle \theta_2^{(1)} (t)\rangle+
\langle {\theta}_2^{(0)} (t)\theta_1^{(1)}(t)\rangle  -\langle {\theta}_2^{(0)} (t) \rangle \langle \theta_1^{(1)}(t) \rangle \right)+
\mathcal{O}(\epsilon^2)
\end{split}
\end{equation}
where $ \theta_k^{(1)}$ is given by eq. (\ref{eq:9}). 
The first terms on the right hand side vanish because  
$\langle \theta_1^{(0)} \theta_2^{(0)} \rangle=
\langle \theta_1^{(0)} \rangle \langle \theta_2^{(0)} \rangle$.
The average is intended as
\begin{equation}
\langle {\theta}_1^{(0)} (t)\theta_2^{(1)}(t)\rangle=
\langle {\theta}_1^{(0)} (t)\theta_2^{(1)}(t)\rangle_{\theta_1^{(0)} ,\theta_2^{(0)} ,..,\bar{I}_k}
\end{equation}
Because $\theta_k^{(0)}$ are uncorrelated, then 
\begin{equation}
\label{eq:9bistheta}
\langle \theta_j^{(0)} \theta_i^{(1)}\rangle_{\theta_i^{(0)}} =
\sum'_{234} T_{i234}\sqrt{\frac{\bar{I}_2\bar{I}_3\bar{I}_4}{\bar{I}_i}} 
\theta_j^{(0)} \frac{ \langle \sin\left(\Delta \theta_{i2}^{34(0)}  \right) - 
 \sin\left(\Delta\bar{\theta}_{i2}^{34}\right)\rangle_{\theta_i^{(0)}}}
{\Delta \bar\Omega_{12}^{34}}\delta_{12}^{34}=0
\end{equation}
The equality to zero is dictated by the fact that $\langle \theta_i^{(1)} \rangle_{\theta_i^{(0)}}=0$.
Similarly, all terms vanish, so that in conclusion
\begin{equation}
\langle \theta_1(t) \theta_2(t) \rangle=\langle \theta_1(t)\rangle \langle \theta_2(t) \rangle
+
\mathcal{O}(\epsilon^2)
\end{equation}

The same procedure can be used to study the correlation of $n$ angles, for $n$ arbitrary-large.
\bibliographystyle{unsrtnat}
\bibliography{references}
\end{document}